# Enhanced TM-Mode 3D Coupled Wave Theory for Photonic Crystal Surface-Emitting Terahertz Quantum Cascade Lasers

MINGXI CHEN,[1,2,*] TSUNG-TSE LIN,[1] LI WANG,[1] HIDEKI HIRAYAMA,[1] AND CHIKO OTANI[1,2,3]

[1]*Terahertz-wave Research Group, RIKEN Center for Advanced Photonics, 519-1399 Aoba, Aramaki, Aoba-ku, Sendai, Miyagi 980-0845, Japan*
[2]*Department of Physics, Graduate School of Science, Tohoku University, 6-3 Aramaki Aza Aoba, Aoba-ku, Sendai 980-8578, Japan*
[3]*otani@riken.jp*
**m.chen@astr.tohoku.ac.jp*

**Abstract:** We propose and develop the transverse magnetic (TM) mode of an enhanced three-dimensional coupled wave theory (TM-mode 3D CWT) to investigate the optical field behavior in photonic crystal surface-emitting terahertz quantum cascade lasers (THz-QCLs). By incorporating an effective permittivity enhancement (EPE) model, we successfully address the numerical dispersion issues encountered in analytical methods when dealing with metallic waveguide structures. The results demonstrate that the EPE 3D CWT achieves computational accuracy comparable to traditional methods such as finite-difference time-domain (FDTD), while significantly reducing the required time and memory resources to mere tens of minutes. Moreover, this method provides a clear physical insight, revealing the reasons behind the current low extraction efficiency in surface-emitting THz-QCLs. Our study showcases the potential of the EPE 3D CWT as a powerful simulation tool in the research of photonic crystal surface-emitting lasers, offering a new theoretical foundation and optimization direction for future laser designs.

## 1. Introduction

Terahertz quantum cascade lasers (THz-QCLs) have garnered significant attention due to their use of semiconductor superlattice structures in the conduction band, forming a series of potential wells, and achieving 1-5 THz band electrical injection laser radiation through electron transitions between these sublevels [1, 2]. As a compact, highly coherent, and energy-efficient THz source, THz-QCLs have shown immense application potential in various fields such as nondestructive testing, biological research, and high-speed telecommunication [3-6]. Recently, THz-QCLs have successfully achieved a maximum output power exceeding 1 W and an operating temperature of 260 K [7-10], marking a significant step toward practical applications. However, due to the typical thickness of the active region (approximately 10 μm) [11, 12] being much smaller than the wavelength of THz radiation in air (about 100 μm), conventional edge-emitting THz-QCLs face severe diffraction issues, resulting in distorted far-field beam patterns [13, 14]. Moreover, current THz-QCLs struggle to achieve both high brightness and high operating temperatures simultaneously; most THz-QCLs only reach milliwatt-level brightness under near-room temperature conditions [9, 10, 15, 16]. Enhancing the far-field beam quality and radiation power has become a crucial challenge for the practical implementation of THz-QCLs.

Surface-emitting THz-QCLs with guided mode resonance (GMR) photonic crystal structures offer unique advantages in addressing these challenges [17, 18]. Compared to traditional edge-emitting designs, surface-emitting lasers have a larger light extraction area, significantly improving the far-field mode and achieving higher radiation power. Additionally, due to the periodic spatial structure of the GMR photonic crystal, these lasers also achieve better single-mode operation. A method similar to that in [19, 20] is expected to apply to THz-QCLs, where over the past decade, numerous studies have been conducted with the aim of achieving

Secondary email (for backup correspondence): chenmingxi.work@gmail.com

high-power surface-emitting THz-QCLs [21-23]. However, unlike conventional semiconductor lasers, which predominantly operate in the transverse electric (TE) mode, THz-QCLs are inherently constrained to transverse magnetic (TM) polarization due to their quantum cascade nature. While some research has explored TM-mode surface emission, it remains significantly less studied compared to its TE-mode counterpart, leaving critical design challenges unresolved. Furthermore, the sub-wavelength thickness of the active region in THz-QCLs necessitates the use of highly reflective materials to ensure strong optical field confinement. While this approach effectively enhances waveguiding, it also exacerbates the challenge of light extraction. As a result, despite extensive efforts, achieving efficient surface emission in TM-mode THz-QCLs remains an unsolved problem.

We believe that developing suitable simulation methods to gain a deeper understanding of the optical field behavior in photonic-crystal-structured surface-emitting THz-QCLs is one of the important keys to addressing this challenge. Traditional numerical simulation methods, such as the finite-difference time-domain (FDTD) and finite element method (FEM) [24, 25], are widely used due to their reliability. However, their high computational cost makes it difficult to efficiently explore a broad range of design configurations. More critically, these methods primarily provide numerical solutions without offering deep physical insights into the mechanisms limiting the extraction efficiency of surface-emitting THz-QCLs.

Recently, the three-dimensional coupled wave theory (3D CWT) has emerged as a powerful analytical approach in the infrared regime, particularly for surface-emitting laser diodes and QCLs [26-29]. By providing direct access to key device characteristics, including operating frequency, electromagnetic field intensity, and surface losses, 3D CWT enables a more efficient and insightful investigation of surface-emitting laser structures. Although 3D CWT has been successfully applied in the infrared range for TE-polarized surface-emitting lasers, its direct application to TM-mode THz-QCLs remains challenging. As we will show, modifications are required to accurately model the strong optical confinement and metal waveguide structures in THz-QCLs.

We plan to utilize this powerful simulation method to analyze photonic-crystal-structured surface-emitting THz-QCLs. However, in THz-QCLs, to maintain optical field confinement, metal waveguides must be introduced. The extreme permittivities of these materials present significant numerical dispersion issues for current 3D CWT calculations, similar to those encountered in FDTD [30]. To fully leverage this powerful analytical model and achieve high radiation power in photonic-crystal-structured surface-emitting THz-QCLs, this study introduces the effective permittivity enhancement (EPE) model to extend the 3D CWT for TM polarization modes, making accurate simulation of THz-QCLs with metal waveguides possible. By comparing with the results of the FDTD method, we have verified the accuracy of this method and, for the first time, successfully revealed the reasons for the low extraction efficiency of current surface-emitting THz-QCLs compared to surface-emitting lasers in other bands using this analytical approach. Through this research, we also demonstrate the potential of the EPE-enhanced TM-mode 3D CWT as an effective simulation tool for the study of photonic crystal surface-emitting lasers, including surface-emitting THz-QCLs.

## 2. Coupled Wave Theory Model

A typical photonic-crystal-structured surface emitting THz-QCLs structure [23, 31, 32] is shown in Figure 1. The optical field between waveguides exhibits different guided mode vertical distributions in the metal top layer and air node regions, resulting in different effective permittivities in various regions [32], as shown in Figure 3. By controlling the periodic distribution of the effective permittivities, photonic-crystal-structured surface emitting THz-QCLs can achieve efficient optical field control. Therefore, the kind of THz-QCLs are typical GMR photonic crystal devices. For GMR photonic crystals, when faced with large air nodes, strong permittivity contrasts $\Delta\varepsilon$ ($\sim 10^1$), or complex photonic crystal structures, it has been reported that iterative optimization of the coupled wave theory is useful to achieve sufficiently high simulation accuracy [33]. However, relying solely on the self-consistent iteration is insufficient for efficiently modeling surface-emitting photonic crystal THz-QCLs. The fundamental challenge lies in the theoretical limitations of the CWT method and the extreme permittivity characteristics of metallic materials in the THz frequency range.

The CWT method is fundamentally based on selecting critical basis modes (known as fundamental waves) and constructing their superpositions to describe the primary features of the electromagnetic field. This approach is conceptually similar to perturbation analysis. CWT, with its finite basis expansion, with conventional semiconductor devices, usually the initial solution obtained is of high quality ($|\omega_{error}| < \omega_{true}$). Under such conditions, the self-consistent iteration method can rapidly converge to the physically accurate solution. However, when systems exhibit extreme permittivity contrasts, the initial solution generated by CWT may become severely ill-conditioned ($|\omega_{error}| \gg \omega_{true}$), rendering it an unreliable starting point for the iterative process. This limitation significantly undermines the applicability of CWT in complex systems.

In THz-QCLs, metallic waveguides play an indispensable role in maintaining the confinement factor of the optical field. However, the permittivity of metals in the terahertz frequency range differs by approximately three orders of magnitude (~$10^2$ to $10^5$) from that of semiconductors. Furthermore, the real part of the metal's permittivity is negative, indicating strong reflection and absorption of electromagnetic fields. This extreme contrast leads to abrupt variations in the optical field distribution at metal-semiconductor interfaces, which, in turn, induces ringing artifacts and severe numerical dispersion (as shown in Fig. 2.B). These issues are particularly pronounced for CWT methods lacking support for high-order basis modes, further limiting their applicability in high-precision simulations.

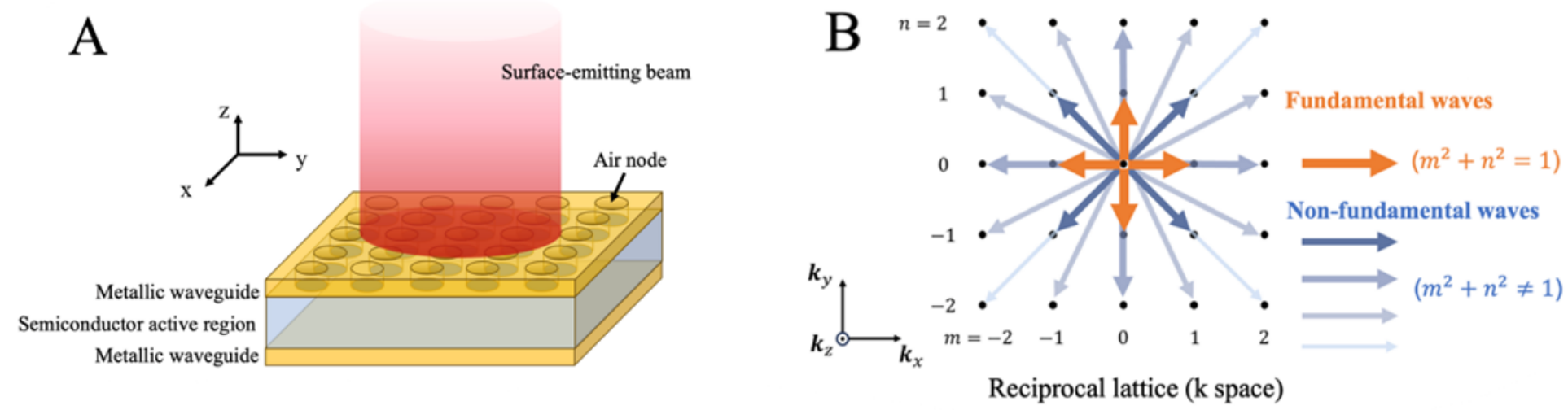

Fig. 1. A. The schematic depicts a typical structure of surface-emitting THz QCL with a two-dimensional GMR photonic crystal waveguide. Based on this structure, we will construct a CWT model. Here, we adopt a square lattice and neglect the ultrathin (~50 nm) heavily doped contact layer between the metal and the active region, which has a permittivity close to that of the metal. Assuming the device is sufficiently large to disregard the side boundary conditions, we can consider the device to be infinitely extended in the xy-plane. B. The reciprocal lattice corresponding to the photonic crystal, along with the wave vectors of the monochromatic Bloch waves. m, n are wavenumbers of each wave.

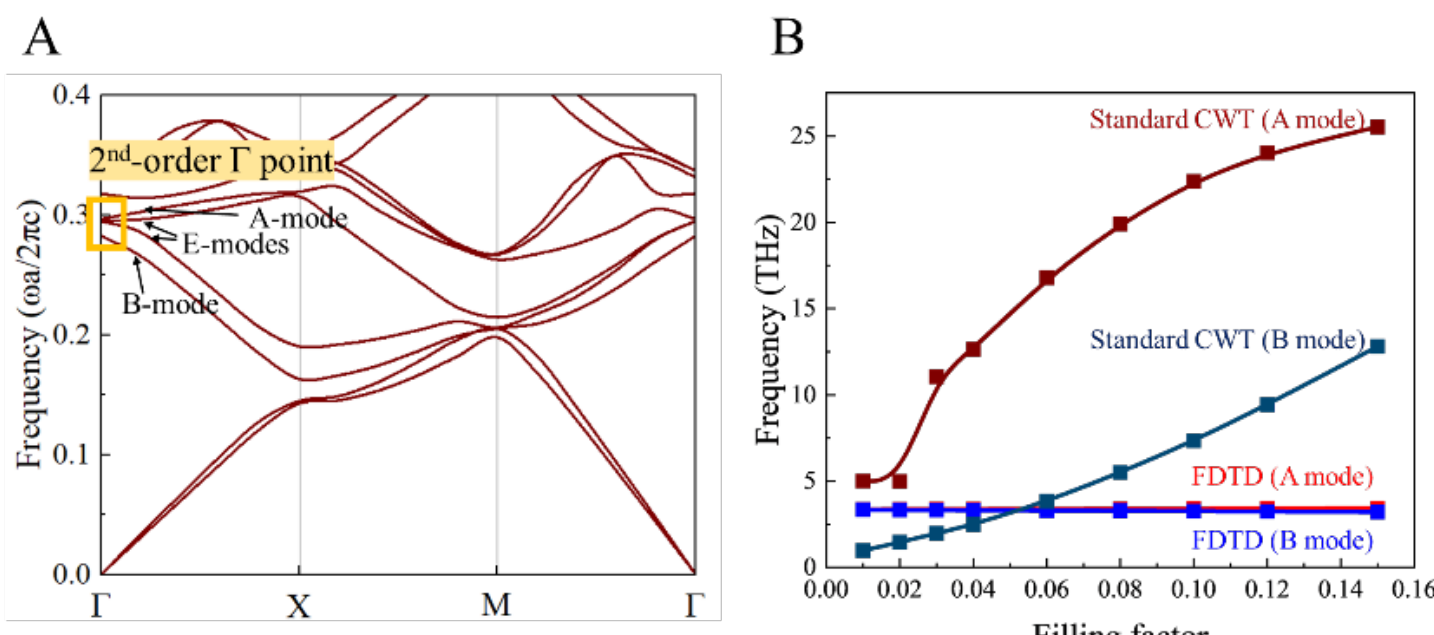

Fig. 2. A. We will focus on establishing a CWT model at the 2nd-order Γ point, as this point yields the highest surface emission efficiency. There are four supported sub-modes here, designated as A, B, and E (E1, E2), which will be discussed in detail later. This band diagram is obtained using two-dimensional plane wave expansion method (2D

PWEM), and as we will see later, it exhibits some discrepancies compared to fully three-dimensional calculations. B. We observe that, due to the influence of the metal, using the standard CWT model alone can result in significant deviations from the relatively accurate values provided by FDTD simulations.

To bridge this significant numerical difference and extend the applicability of CWT to include plasmonic waveguides containing metals or heavily doped semiconductors, we introduce an EPE algorithm. This algorithm avoids direct calculation of the metals, simplifying the problem to an equivalent model, thereby significantly reducing the permittivity contrast encountered by CWT when modeling THz-QCLs to a level comparable to that of conventional semiconductor devices. Thus, achieving high precision similar to that of coupled wave calculations conducted entirely on semiconductor devices. For clarity, the following sections will provide a step-by-step introduction to the standard TM-mode 3D CWT. Descriptions of the self-consistent iteration method and EPE will follow. To ensure clear and coherent derivation, the physical meanings of most symbols and notations required in the subsequent discussions are uniformly provided in Table 1.

**Table 1 Symbols used in derivation and their meanings.**

| Symbol | Physical quantity |
|---|---|
| H | Magnetic field intensity |
| E | Electric field intensity |
| $\varepsilon(\boldsymbol{r})$ | Permittivity (Permittivity) at space coordinates $\boldsymbol{r}$ |
| $k$ | Wave number |
| $k_0$ | Resonant mode wave number |
| $\omega$ | Angular frequency |
| $\omega_0$ | Resonant mode angular frequency |
| $\beta_0$ | Reciprocal unit cell size $= \frac{2\pi}{a(unit\ size)}$ |
| $m, n, m', n'$ | Fourier expansion indices |
| $M, N$ | Fourier expansion indices of fundamental waves, $M^2 + N^2 = 1$ for 2nd-Γ point |
| $\kappa_{mn}$ | Fourier transform of the inverse permittivity |
| | $\varepsilon^{-1}(\boldsymbol{r}) = \sum_{mn} \kappa_{mn}(z) e^{-i\beta_0(mx+ny)}$ |
| Θ(z) | Vertical field distribution of fundamental waves |
| $A_{MN}$ | Complex amplitude of fundamental waves |
| $\widehat{K}_{mn}^{MN}$ | Dispersion relation operator $= \kappa_{(M-m)(N-n)} \frac{\partial^2}{\partial z^2} + \left(\frac{\partial}{\partial z} \kappa_{(M-m)(N-n)}\right) \frac{\partial}{\partial z}$ |
| $\hat{J}$ | Dispersion relation operator $= \widehat{K}_{mn}^{mn} + k_0^2 = \kappa_{00} \frac{\partial^2}{\partial z^2} + \left(\frac{\partial}{\partial z} \kappa_{00}\right) \frac{\partial}{\partial z} + k_0^2$ |

## *2.1 Standard TM-mode 3D CWT Mode*

For the TM-mode optical field, the electric and magnetic fields can be expressed as $\boldsymbol{E} = (0,0,E_z)$ and $\boldsymbol{H} = \left(H_x, H_y, 0\right)$ [29], respectively. To retain the polarization information of these fields, we choose the dispersion equation represented by the magnetic field $\boldsymbol{H}$ as the starting point for our discussion, from the Faraday's law of electromagnetic induction and Ampère's law with Maxwell's correction, the master equation is given by

$$\nabla \times \left[\frac{1}{\varepsilon(\boldsymbol{r})} \nabla \times \boldsymbol{H}(\boldsymbol{r})\right] = k^2 \boldsymbol{H}(\boldsymbol{r}) \qquad (1)$$

We select a laser with a multilayer thin-film structure, as shown in Figure 1, as the subject of our study. The two-dimensional photonic crystal is located in the x-y plane, while the multilayer structure extends along the z-axis. This is a typical structure for many surface-emitting lasers, including surface-emitting THz-QCLs. Considering the desire to reduce lateral mirror losses, improve surface emission brightness, and achieve smaller beam divergence angles and single-mode operation, photonic crystal surface-emitting lasers are typically designed with larger dimensions and more repeating units in the x-y direction. Therefore, assuming the device is infinite in the x-y direction with countless repeating periods and negligible horizontal losses is a reasonable approximation. Based on this discussion, we can fully utilize the periodicity in the x-y plane and apply the Bloch wave expansion to the above equation (i.e., Fourier transform). Thus, the magnetic field can be expanded as:

$$\boldsymbol{H}(\boldsymbol{r}) = \sum_{mn} e^{-i\beta_0(mx+ny)} \cdot \boldsymbol{H}_{mn}(z) = \sum_{mn} e^{-i\beta_0(mx+ny)} \cdot \left(H_{x,mn}, H_{y,mn}, 0\right) \tag{2}$$

Similarly, the inverse of the dielectric function can be expanded as:

$$\varepsilon^{-1}(\boldsymbol{r}) = \sum_{mn} \kappa_{mn}(z) e^{-i\beta_0(mx+ny)} \tag{3}$$

Substituting the above two equations into the dispersion equation, we obtain a set of equations describing the coupling relationships between various monochromatic frequency components in the x, y, and z directions:

$$\begin{gathered}\left(\hat{J} - \beta_0^2\kappa_{00}n'^2\right)H_{x,m'n'} + \beta_0^2\kappa_{00}m'n'H_{y,m'n'} \\ = \sum_{\substack{m\neq m' \\ n\neq n'}} \left[\left(-\widehat{K}_{mn}^{m'n'} + \beta_0^2\kappa_{(m'-m)(n'-n)}nn'\right)H_{x,mn} - \beta_0^2\kappa_{(m'-m)(n'-n)}mn'H_{y,mn}\right]\end{gathered} \tag{4}$$

$$\begin{gathered}\left(\hat{J} - \beta_0^2\kappa_{00}m'^2\right)H_{y,m'n'} + \beta_0^2\kappa_{00}m'n'H_{x,m'n'} \\ = \sum_{\substack{m\neq m' \\ n\neq n'}} \left[\left(-\widehat{K}_{mn}^{m'n'} + \beta_0^2\kappa_{(m'-m)(n'-n)}mm'\right)H_{y,mn} - \beta_0^2\kappa_{(m'-m)(n'-n)}m'nH_{x,mn}\right]\end{gathered} \tag{5}$$

$$\sum_{m,n} \kappa_{(m'-m)(n'-n)} \frac{\partial}{\partial z}\left(m'H_{x,mn} + n'H_{y,mn}\right) = 0 \tag{6}$$

We can categorize the various monochromatic waves present in the device based on the plane wave numbers into three groups: fundamental waves (FW), surface radiated waves (SR), and higher-order waves (HO). The wave number of the fundamental waves is $\beta_{MN} = \sqrt{M^2 + N^2}\beta_0$, satisfying $\beta_{MN} \approx \omega_{act}/c$, where $\omega_{act}$ is the angular frequency of the radiation from the active region. Since the wave numbers of the fundamental waves are very close to the wave numbers of the radiation from the active layer, they will have the largest amplitude, dominating the optical field properties of the entire device. On the other hand, the longitudinal plane wave number of the surface radiated waves is zero, meaning they can only propagate along the z direction, representing the components of surface radiation. Higher-order waves refer to other monochromatic wave components that are neither fundamental nor radiated waves. Intuitively, the closer the wave numbers of the fundamental and radiated waves, the higher the efficiency of generating surface radiation through coupling. Evidently, near the 2nd-Γ point of the photonic crystal band diagram, photonic crystal surface-emitting lasers exhibit the best light extraction efficiency. Therefore, in subsequent discussions, we will also consider the waves with $M^2 + N^2 = 1$ as the fundamental waves.

In the resonant modes near the 2nd-Γ point, the fundamental waves can be expressed as the product of the same longitudinal optical field distribution $\Theta(z)$ and their respective complex amplitudes, specifically:

$$\begin{gathered}H_{x,10} = 0, \quad H_{y,10} = A_{10}\Theta(z) \\ H_{x,-10} = 0, \quad H_{y,-10} = A_{-10}\Theta(z) \\ H_{x,10} = A_{01}\Theta(z), \quad H_{y,10} = 0 \\ H_{x,-10} = A_{0-1}\Theta(z), \quad H_{y,10} = 0\end{gathered} \tag{7}$$

Here, $\Theta(z)$ can be obtained through the Transfer Matrix Method (TMM) on the waveguide with an averaged permittivity approximation (See. Supplemental Document), satisfying the following equation:

$$\left(\kappa_{00} \cdot \frac{\partial^2}{\partial z^2} + \frac{\partial}{\partial z} \kappa_{00} \cdot \frac{\partial}{\partial z} + k_0^2 - \beta_0^2 \kappa_{00}\right) \Theta(z) = 0 \tag{8}$$

Combining the above equations, we can obtain the following relationship:

$$\begin{aligned} k^2 A_{MN} &= (k_0^2 + \delta_{FW} + \delta_{SR} + \delta_{HO}) A_{MN} \\ &= k_0^2 A_{MN} + \sum_{\substack{m \neq M \\ n \neq N}} \int dz \left\{ \Theta^* \cdot \begin{bmatrix} -\widehat{K}_{mn}^{MN} \left(N^2 H_{x,mn} + M^2 H_{y,mn}\right) \\ -\beta_0^2 \kappa_{(M-m)(N-n)} (M^3 - N^3)(n H_{x,mn} - m H_{y,mn}) \end{bmatrix} \right\} \end{aligned} \tag{9}$$

Note that here we have used the assumption $M^2 + N^2 = 1$. This equation describes a clear physical image: although the fundamental waves dominate the device behavior (the $k_0^2 A_{MN}$ term on the right side of the equation), due to the presence of relatively weak non-fundamental waves, their coupling with the fundamental waves (the summation term on the right side of the equation) causes a shift in the working frequency $k$ of the photonic crystal relative to the "clean" resonant mode $k_0$, with the magnitude of the shift attributed to $\delta_{FW}$ (shift due to coupling with fundamental waves), $\delta_{SR}$ (shift due to coupling with surface radiated waves), and $\delta_{HO}$ (shift due to coupling with higher-order waves).

The above discussion reveals a physical picture: in photonic crystal surface-emitting lasers, the intensities of monochromatic waves are not entirely independent due to their coupling relationships. Furthermore, the fundamental waves will dominate the electromagnetic field behavior throughout the device. Therefore, it naturally follows to expand the monochromatic waves using the four fundamental waves as a basis. This corresponds to a natural and clear physical process: in photonic crystal surface-emitting lasers, the fundamental waves are initially generated by the radiation from the active region, while other monochromatic wave components result from the coupling of fundamental waves as they propagate through the device:

$$H_{mn,\alpha} = {T_{mn,\alpha}^{MN}}^{(1)} \cdot A_{MN} + {T_{mn,\alpha}^{MN,M'N'}}^{(2)} \cdot A_{MN} \cdot A_{M'N'} + O^{(3)}(A_{MN}) \quad (\alpha = x, y) \tag{10}$$

For simplicity, Einstein summation notation is used here, with $T$ being the expansion coefficient, and $O^{(n)}$representing the n$^{th}$-order small quantity. In the first-order approximation, Equation 9 can be transformed into an eigenvalue problem:

$$k^2 \boldsymbol{V} = C\boldsymbol{V} = (k_0^2 + C_{FW} + C_{SR} + C_{HO}) \boldsymbol{V} \tag{11}$$

where, $\boldsymbol{V} = (A_{10}, A_{-10}, A_{01}, A_{0-1})$ represents the complex amplitude vector of the fundamental waves. The specific form and derivation of the operator $C$ are provided in Supplemental Document.

### *2.2 Effective Permittivity Enhancement Model*

In photonic crystal surface-emitting lasers, the presence of strong permittivity contrasts or complex photonic crystal structures poses significant challenges for standard CWT. This is primarily because high-reflectivity interfaces or intricate structures necessitate sharper high-order components to accurately describe the more rapid spatial variations of the electromagnetic field, as we have explained. However, since CWT typically relies on a finite number of Bloch waves and primarily focuses on the fundamental modes, it fails to accurately model devices incorporating metallic photonic crystal structures, such as photonic-crystal-structured surface-emitting THz-QCLs.

To address this limitation, a self-consistent iterative approach has been proposed in prior studies to refine traditional CWT calculations. The effectiveness of this method stems from the following mathematical framework: in conventional TM-mode CWT, the amplitudes of monochromatic waves $H_{mn}$ are derived from the first-order expansion of the fundamental mode amplitudes $A_{MN}$. Consequently, in the dispersion equation, the coupling between each monochromatic wave and the fundamental mode is based on the first-order interaction of $A_{MN}$, which can be regarded as a second-order perturbation expansion:

$$k^2 \approx k_0^2 + \delta k_1^2 O^{(1)}(A_{MN}) + \delta k_2^2 O^{(2)}(A_{MN})$$

Evidently, the self-consistent iterative CWT solution process incorporates higher-order $\delta k_n^2 O^{(n)}(A_{MN})$ terms, and the iteration can be repeated multiple times until the desired accuracy is achieved. This method circumvents the computational challenges associated with directly solving for the amplitudes of hundreds of plane waves, forming a semi-analytical approach. However, if the actual solution $k_{true}^2$ deviates significantly from the initial (resonant mode) approximation $k_0^2$ beyond its intrinsic scale, the fundamental assumptions of perturbation theory collapse, rendering self-consistent-iteration-based corrections to CWT inadequate for precise calculations.

To overcome this issue, a method bridging the gap between $k_{true}^2$ and $k_0^2$ is required. Fortunately, although the unique electromagnetic behavior introduced by metals complicates direct CWT analysis, in practical applications, the electromagnetic field penetration into metals is extremely weak. This implies that the impact of metal is predominantly confined to its exterior regions. We leverage this insight by employing an effective permittivity approach, a concept widely utilized in GMR photonic crystals.

In the case of THz-QCLs, although the TM-mode optical field primarily propagates within the GaAs active region, the significant permittivity contrast at the metal-semiconductor and GaAs-air interfaces results in distinct guided-mode characteristics in different regions. Specifically, the wavevector of the guided modes varies across different regions at the same frequency, effectively causing the optical field to "experience" different dielectric environments, which manifests as a spatial variation in the effective permittivity. As illustrated in Fig. 4, we leverage this concept to construct an equivalent computational model, referred to as the EPE CWT, which circumvents the numerical complexities associated with directly solving the electromagnetic behavior in metallic photonic crystals.

The effective permittivity is determined based on the guided-mode characteristics obtained via the TMM. Through the EPE CWT framework, we can effectively model the influence of metallic waveguides on the optical field without explicitly computing the contributions of infinitely high-frequency Bloch components, while preserving a clear physical interpretation. A key advantage of this approach is that by employing EPE and its equivalent structure, we rationally adjust the permittivity contrast to a level comparable to that of conventional all-semiconductor devices, thereby significantly reducing numerical instabilities and enabling more accurate computational predictions. Further details on this methodology are provided in Supplemental Document A.

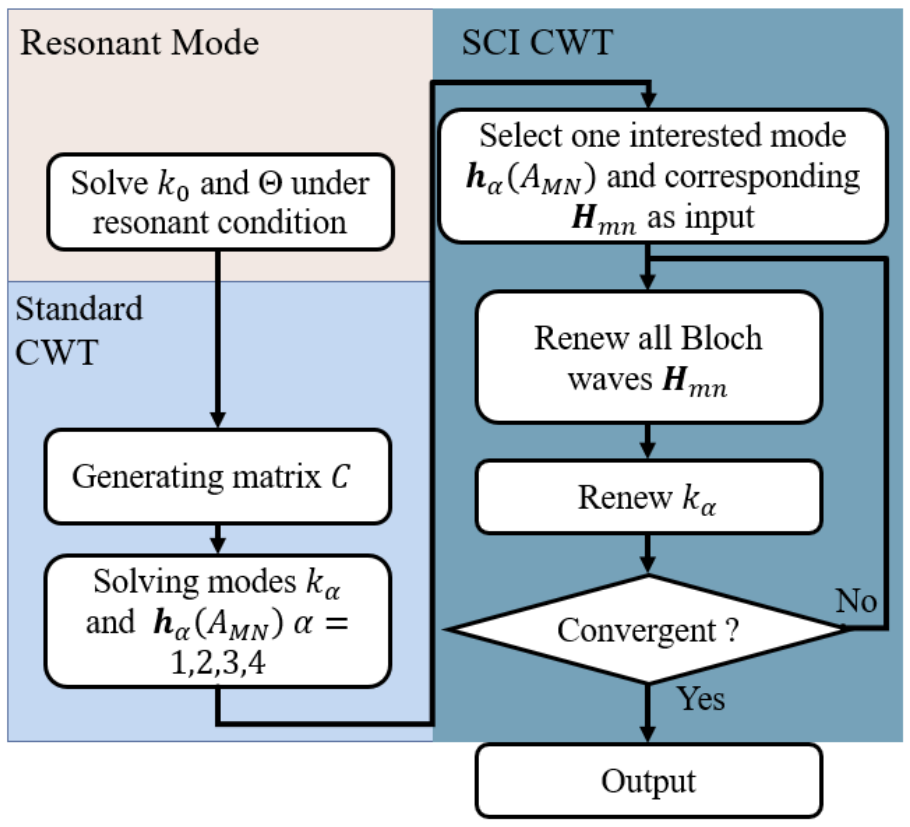

Fig. 3. Depicts the flowchart of TM mode CWT calculations. The program begins in the upper left corner and concludes in the lower right corner. It starts from zeroth order mean field effects and proceeds through resonant mode calculations, Standard CWT calculations, and SCI-CWT calculations. EPE will be introduced from very first step, if it is needed. It is an iterative process from first-order perturbation to low-contrast approximation and finally to second-order-accuracy perturbation, involving higher-order-accuracy perturbations that provide quasi-exact solutions.

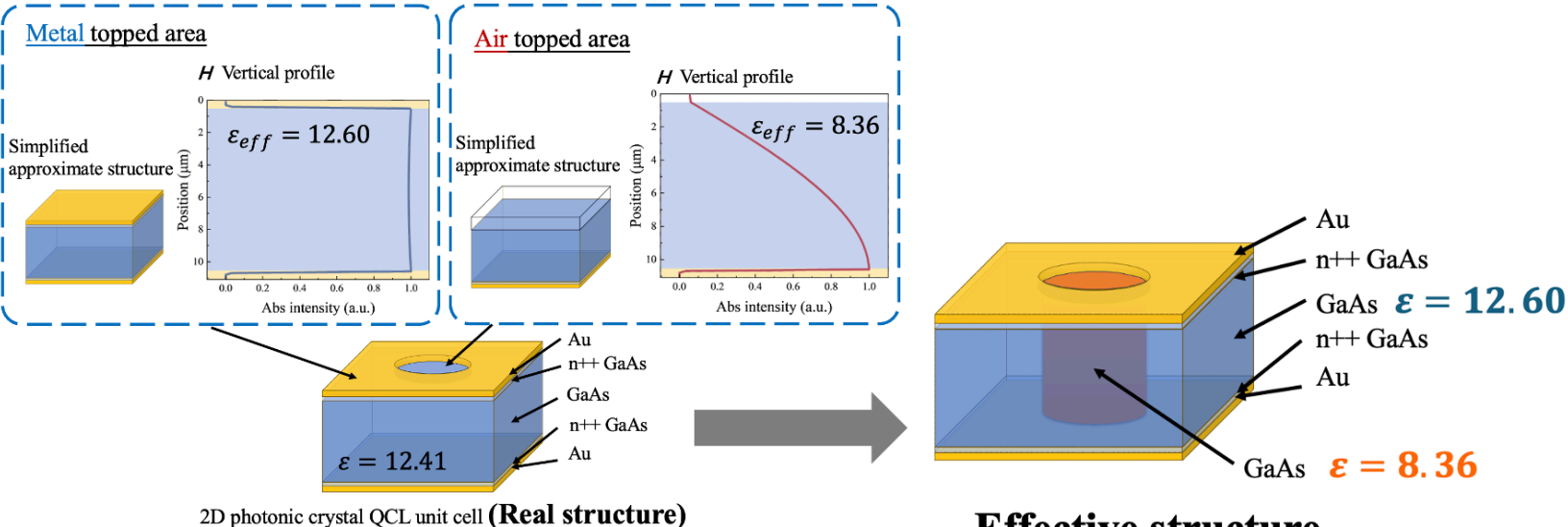

Fig. 4. The conceptual diagram of the EPE CWT. By utilizing the effective permittivity obtained from different guided modes, we can transform the actual homogeneous GaAs active layer into an effective structure with varying permittivity. In our calculations, the imaginary parts of the effective permittivities are small (~0.02). Therefore, we considered only the real part of the dielectric constant to simplify the computation. This approach allows us to incorporate the influence of the metal within the semiconductor layers, thus avoiding direct calculations involving the metal. Given the relatively mild differences in permittivity between semiconductors in the effective structure, we can expect high accuracy and rapid convergence in the calculations.

## 3. Computational Results and Discussion

In this section, we present the computational results obtained from the EPE enhanced TM-mode 3D CWT model, specifically for a typical surface-emitting THz QCL structure containing metallic photonic crystal waveguides. In the calculations, we adopted a plane-wave truncation condition of $Max(m, n) = 10$, resulting in 21×21 plane waves included in the computation. Due to the excellent symmetry of the validation structure and the significant reduction in permittivity differences achieved by EPE, the results fully converged after just one iteration.

For comparison, we also utilized the traditional numerical method FDTD for simulating the same structure using Opti Wave FDTD software. Due to the significant contrast between the semiconductor layers and the metallic material with a complex permittivity having a negative real part, the FDTD calculations tend to be unstable. Thus, we approximated the metal and the 50 nm thick contact layer as ideal conductors. Both the TM-mode 3D CWT and FDTD simulations employed identical periodic boundary conditions, and the actual calculation region size was one unit cell.

It is important to note that the CWT calculations here are based on an equivalent structure derived using the EPE. In the vertical direction, the optical field distribution is first calculated using the actual permittivities of the materials to obtain an averaged effect. This is followed by simulating the coupling within the active region using the equivalent permittivity, which ultimately yields the computational results. In contrast, FDTD, as the primary validation method, directly simulates a structure as close to the actual device as possible, without any additional processing for the active region (although an ideal conductor approximation was applied to the metal and contact layer as we mentioned above).

In our simulation of surface-emitting THz-QCL structures, we observe exceptionally strong optical field confinement and minimal energy loss, resulting in an ultrahigh cavity Q-factor. However, due to the steep spatial variations of the optical field, the finite-difference time-domain (FDTD) method tends to generate ringing artifacts, which gradually accumulate and eventually dominate the numerical solution. As a result, unlike conventional semiconductor waveguides with lower Q-factors, FDTD struggles to achieve a fully relaxed steady-state solution within a feasible computational time.

Although we continue to use Fast Fourier Transform (FFT) and Discrete Fourier Transform (DFT) to extract the eigenfrequencies of the current optical field, the full width at half maximum (FWHM) obtained from these transforms cannot be directly employed to estimate waveguide loss, i.e., surface emission efficiency. Notably, at the 2nd-order gamma point, the three excited sub-modes are mutually orthogonal, allowing us to directly fit the time-amplitude evolution sequence using multiple independent monochromatic modes. A comparison between the FFT-extracted frequencies and the fitted sub-mode frequencies shows an almost perfect agreement. The calculated surface loss is presented later in Figure 7.

Additionally, we introduced the semi-analytical PWEM for comparison. Considering the structure is not periodic along the z-direction, performing full 3D PWEM calculations requires additional considerations. Here, we used its 2D form (2D PWEM) and applied the same effective permittivity approximation as in the CWT, treating the equivalent structure in PWEM calculations as the holes ($\varepsilon_{eff} \approx 8.375$) and the slab ($\varepsilon_{eff} \approx 12.602$). This is a reasonable and validated approximation for semiconductor devices.

First, we observed that all three methods correctly identified four supported sub-modes at the 2nd-Γ point: two non-degenerate modes A and B, and two degenerate modes E1 and E2. The in-plane electric field distributions obtained from the CWT calculations are shown in Figure 5, with the patterns dependent on the symmetry of the unit cell and air nodes. The results are consistent with those from FDTD and PWEM.

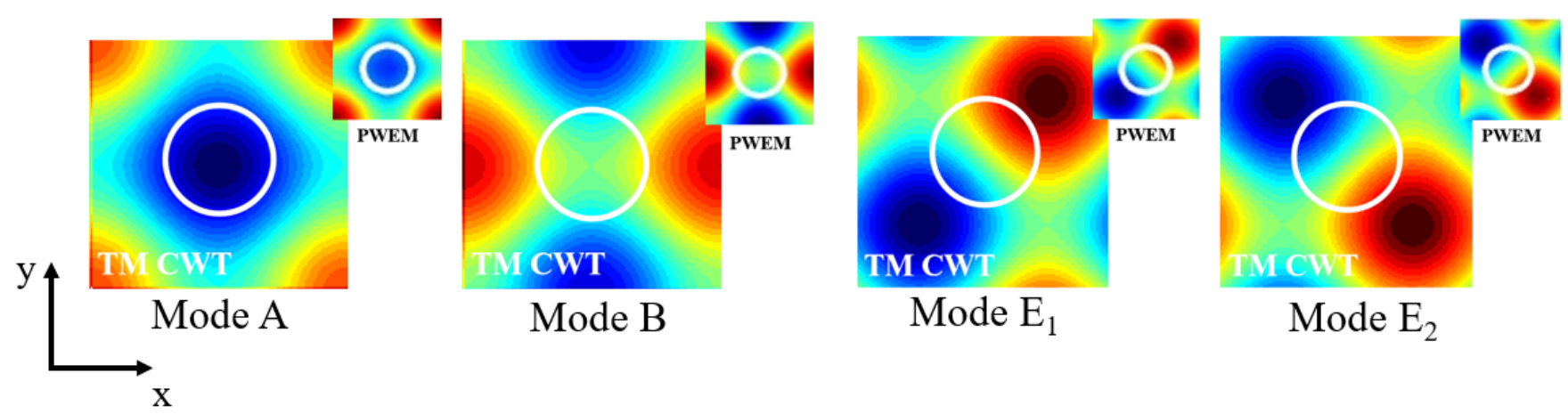

Fig. 5. In-plane Ez field patterns of antisymmetric (A, B) and symmetric (E) modes simulated by CWT and PWEM.

After confirming that the new CWT method correctly identified the four sub-modes, we proceeded to calculate the two primary factors of concern in photonic crystal surface-emitting THz QCL designs: operating frequency and surface emission efficiency. Firstly, regarding frequency, we varied the air node filling factor (filling factor = Air node area/unit cell area) between 0.01 and 0.15. We used a smaller upper limit for the air node size compared to infrared all-semiconductor surface-emitting lasers because excessive air node sizes have been reported to lead to device failure [32]. This is understandable as QCLs require a high optical confinement factor to achieve emission, and the process of creating air nodes inevitably damages the microstructure of QCLs. The computational results are shown in Figures 6 and 7.

The results demonstrate excellent consistency between the new CWT method and the FDTD method across the computed parameter range for all three modes, with a maximum frequency deviation of only about 2% (~0.07 THz). Moreover, the EPE enhanced TM-mode 3D CWT method showed significant accuracy advantages over the traditional semi-analytical 2D PWEM method. This is because the TM-mode 3D CWT considers the behavior of electromagnetic waves along the z-direction, which is crucial for THz QCLs with high reflectivity waveguides. Additionally, we also have presented the results from the standard TM-mode 3D CWT without the EPE enhanced in Figure 2B. As shown, the computation results diverged significantly due to the effects of the metal, further proving the necessity and correctness of introducing the new EPE techniques when dealing with extreme permittivity materials.

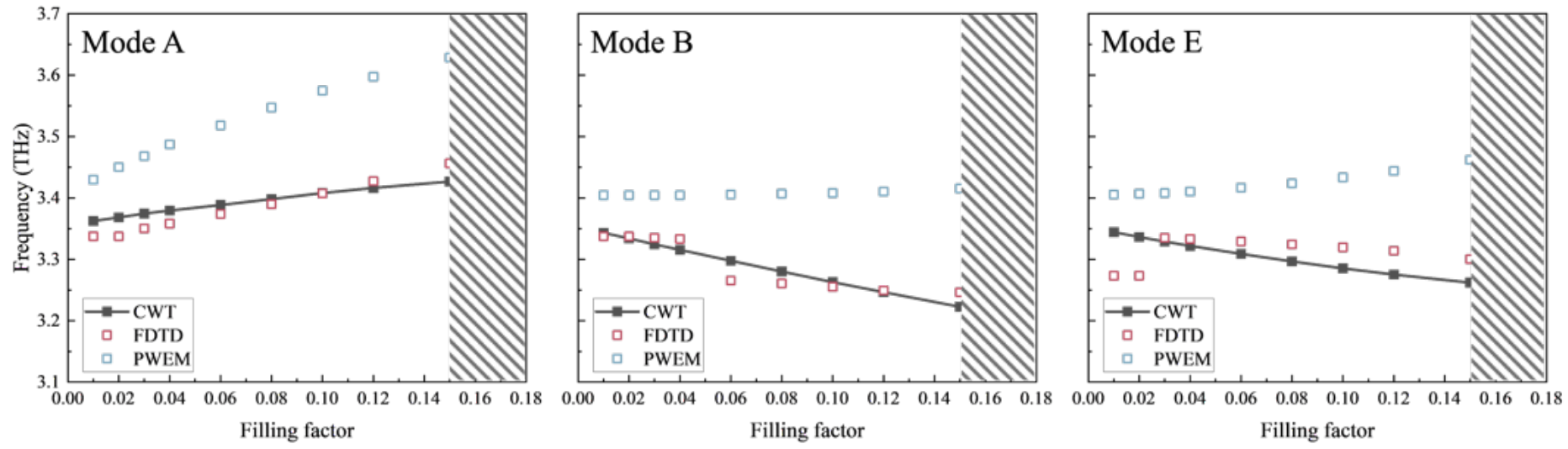

Fig. 6. Frequencies supported by photonic crystals predicted by CWT, FDTD, and PWEM for modes A, B, and E, as a function of air node filling factor (0.01–0.15). Filling factors above 0.15 are excluded due to potential device damage[32]. For THz QCLs, the sub-wavelength active region requires a waveguide with high reflectivity to confine sufficient optical fields and achieve lasing. Larger air holes reduce confinement, compromising functionality. Moreover, increasing air node size causes significant deviation in electric field modes, particularly for mode E, where CWT predictions diverge from FDTD results. Thus, the upper calculation limit is set at 0.15.

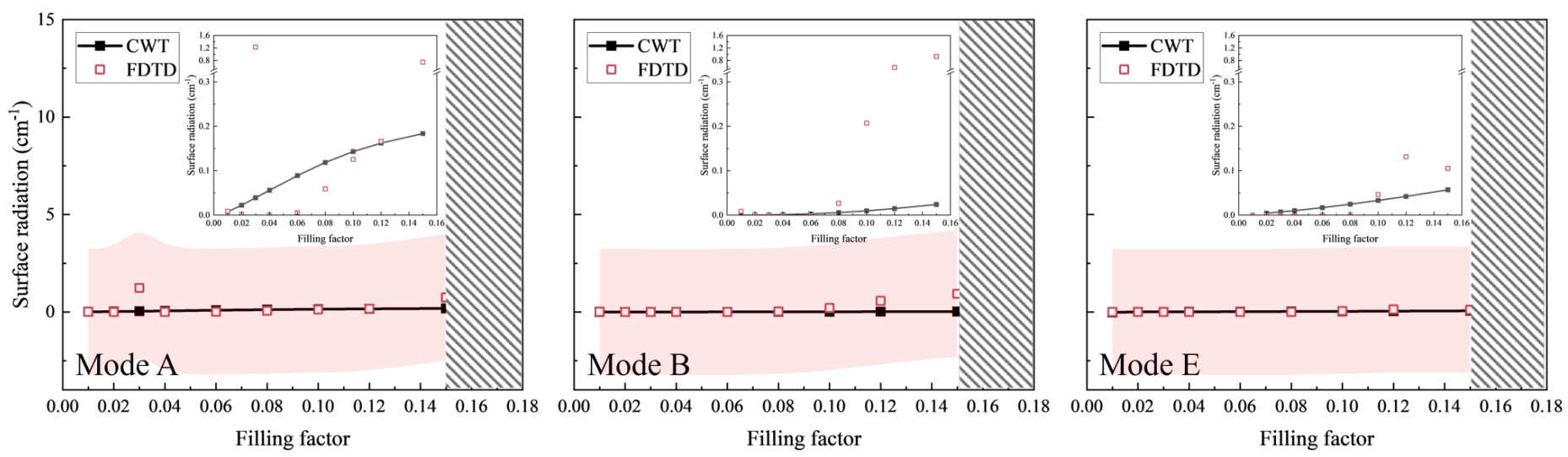

Fig. 7. Comparison of the surface radiation efficiency of THz QCLs simulated using CWT and FDTD methods. Notably, PWEM cannot predict radiation efficiency in the vertical (z-axis) direction, highlighting a key advantage of the semi-analytical CWT method. The main figure presents the overall radiation efficiency, while the inset zooms in on the detailed numerical values. The light red shaded area represents the estimated range of surface radiation efficiency based on FDTD numerical error analysis (see Supplementary Material for details on error estimation). Within the investigated parameter range, both CWT and FDTD simulations yield radiation efficiencies that are extremely close to zero and significantly lower than the estimated numerical error of FDTD. Furthermore, due to the high complex permittivities of metallic (e.g., Au: $\varepsilon \approx -183000 + 167000i$ at 3.5 THz) and heavily doped semiconductor layers (e.g., n++ GaAs: $\varepsilon \approx -394 + 185i$ at 3.5 THz), the propagation of electromagnetic waves experiences additional phase shifts. This effect results in Mode A exhibiting higher radiation efficiency than Mode E, deviating significantly from the behavior observed in ideal purely real-permittivity devices. Nevertheless, none of the three modes demonstrates sufficient surface radiation efficiency to enable effective surface emission.

Another critical aspect of photonic crystal surface-emitting THz QCL performance is its surface emission efficiency, where the 3D CWT method shows a distinct advantage over traditional analytical methods. Since we run full 3D simulations without requiring periodic structures along the z-direction, this method allows us to calculate energy emission efficiency along the z-direction. In contrast, PWEM requires periodicity in all calculation directions (or the introduction of superlattice structures or absorbing layers), making it challenging to compute surface emission efficiency using PWEM.

This time, we observed that unlike all-semiconductor photonic crystal surface-emitting lasers, both the FDTD and TM-mode 3D CWT methods displayed surface emission efficiencies very close to 0 (< 0.2 $cm^{-1}$) regardless of the air node size variation. The good news is that once again, we verified the accuracy of the EPE TM-mode 3D CWT results using the traditional and reliable FDTD method. However, unfortunately, the results indicate an inherent deficiency in surface emission performance for 2D photonic crystal waveguide structures utilizing the GMR effect. Considering that its surface radiation loss is lower than the typical lateral mirror loss of QCLs (~1-20 $cm^{-1}$), surface-emitting QCLs without special treatment to obtain unique boundary conditions are prone to revert to lateral emission behavior. This conclusion aligns with earlier studies on surface-emitting QCLs and our group's experimental results. To date, all highly efficient surface-emitting THz QCLs have employed specialized structures to achieve high emission efficiency [22, 32].

As a semi-analytical method, the TM-mode 3D CWT allows us to derive a clear physical image from the computation formulas, enabling us to directly understand the reasons behind the low emission efficiency of photonic crystal surface-emitting THz QCLs. This is something that traditional FDTD methods cannot achieve. It may be the lack of intuitive insights into the problem's nature that has prevented the realization of high-power surface-emitting THz QCL structures similar to those in other wavelength ranges over the past 20 years, despite many efforts and significant discoveries.

By directly extracting the calculation formula related to surface emission, we see that:

$$I_{surf} \approx \int \left[ \Theta^{*}(z) \cdot \kappa(z) \frac{\partial^2}{\partial z^2} \left\{ \int dz' \left[ G(z,z') \kappa(z') \frac{\partial^2}{\partial z^2} \Theta(z') \right] \right\} \right] dz \tag{12}$$

The surface emission capability of the laser we are considering fundamentally arises from the uncertainty in the photon's momentum, specifically diffraction effects, represented by the second derivative term $\frac{\partial^2}{\partial z^2} \Theta(z)$. It is evident that plane waves propagating strictly in the horizontal direction struggle to achieve vertical components through certain sandwich-like GMR structures. A well-designed photonic crystal structure amplifies the z-component caused by diffraction, leading to notable surface emission. However, due to the high reflectivity of both the upper and lower waveguides, the upward and downward propagating components nearly cancel each other out, allowing only electromagnetic waves strictly confined to the horizontal direction, which making it naturally difficult to produce sufficient surface emission components. Note that we used ideal periodic boundary conditions for the calculations, so the finite scale structure may produce slightly higher surface emission efficiency due to imbalances in electromagnetic field distribution, but this does not affect our discussion results. Solutions to improve this issue will be detailed in our subsequent papers.

## 4. Summary

This paper proposes and develops an enhanced TM-mode 3D CWT to address the numerical dispersion issues encountered in traditional analytical methods when dealing with waveguide structures made of metallic or metal-like materials. By incorporating an EPE model, we significantly improved the computational accuracy and efficiency of 3D CWT in simulating photonic crystal surface-emitting THz-QCLs. Compared to traditional numerical simulation methods such as the FDTD method, our enhanced approach not only drastically reduces the computational resources required but also achieves similar accuracy within tens of minutes. In the case studies presented, the EPE and SCI-improved TM-mode 3D CWT showed excellent agreement with the FDTD method in frequency calculations, with a maximum frequency deviation of only about 2%.

Over the past decade, surface-emitting THz-QCLs have faced the challenge of low extraction efficiency. Benefiting from the semi-analytical nature of CWT, the EPE enhanced TM-mode 3D CWT accurately reveals the reasons for the low extraction efficiency in current surface-emitting THz-QCLs: this is primarily due to the high reflectivity of the metallic waveguides, which causes the electromagnetic field intensity in the sub-wavelength thickness of the active layer to be highly symmetrical, leading to the cancellation of the upward and downward propagating components. Consequently, most of the electromagnetic waves strictly propagate along the xy-plane. This discovery provides a new theoretical foundation for understanding and optimizing surface-emitting THz-QCLs. Based on these findings, future work will focus on further optimizing the photonic crystal structures and surface-emitting conditions to achieve more efficient laser designs.

It is noteworthy that the methods presented in this paper are not limited to THz-QCLs but can also be widely applied to other photonic crystal surface-emitting lasers using metallic or metal-like waveguides, demonstrating potential in a broader range of applications. The EPE enhanced 3D CWT, as a powerful simulation tool, not only provides clear physical insights but also significantly improves computational efficiency, making it highly valuable for future laser design and optimization.

**Acknowledgment.** Mingxi Chen thanks the Junior Research Associate (JRA) program at RIKEN for supporting this work.

The authors declare no conflicts of interest.

## References

1. B. S. Williams, "Terahertz quantum-cascade lasers," Nature photonics **1**, 517-525 (2007).
2. H. Hirayama, W. Terashima, T.-T. Lin et al., "Recent progress and future prospects of THz quantum-cascade lasers," Novel In-Plane Semiconductor Lasers XIV **9382**, 157-167 (2015).
3. W. L. Chan, J. Deibel, and D. M. Mittleman, "Imaging with terahertz radiation," Reports on progress in physics **70**, 1325 (2007).
4. P. H. Siegel, "Terahertz technology," IEEE Transactions on microwave theory and techniques **50**, 910-928 (2002).
5. G. Scalari, C. Walther, M. Fischer et al., "THz and sub‐THz quantum cascade lasers," Laser & Photonics Reviews **3**, 45-66 (2009).
6. Y. H. Tao, A. J. Fitzgerald, and V. P. Wallace, "Non-contact, non-destructive testing in various industrial sectors with terahertz technology," Sensors **20**, 712 (2020).
7. T.-T. Lin, L. Wang, K. Wang et al., "Over One Watt Output Power Terahertz Quantum Cascade Lasers by Using High Doping Concentration and Variable Barrier‐Well Height," physica status solidi (RRL)–Rapid Research Letters **16**, 2200033 (2022).
8. L. Li, L. Chen, J. Freeman et al., "Multi‐Watt high‐power THz frequency quantum cascade lasers," Electronics Letters **53**, 799-800 (2017).
9. A. Khalatpour, A. K. Paulsen, C. Deimert et al., "High-power portable terahertz laser systems," Nature Photonics **15**, 16-20 (2021).
10. A. Khalatpour, M. C. Tam, S. J. Addamane et al., "Enhanced operating temperature in terahertz quantum cascade lasers based on direct phonon depopulation," Applied Physics Letters **122** (2023).
11. T. T. Lin, and H. Hirayama, "Variable Barrier height AlGaAs/GaAs quantum cascade laser operating at 3.7 THz," physica status solidi (a) **215**, 1700424 (2018).
12. A. M. Andrews, T. Zederbauer, H. Detz et al., "THz quantum cascade lasers," in *Molecular Beam Epitaxy*(Elsevier, 2018), pp. 597-624.
13. D. Shao, C. Yao, Z. Fu et al., "Terahertz quantum cascade lasers with sampled lateral gratings for single mode operation," Frontiers of Optoelectronics **14**, 94-98 (2021).
14. S. Fathololoumi, E. Dupont, S. Razavipour et al., "Electrically switching transverse modes in high power THz quantum cascade lasers," Optics express **18**, 10036-10048 (2010).
15. T.-T. Lin, W. Terashima, and H. Hirayama, "250 mW output power operation of GaAs-based THz quantum cascade lasers," in *JSAP-OSA Joint Symposia*(Optica Publishing Group2017), p. 7a_A409_406.
16. A. Albo, and Q. Hu, "Investigating temperature degradation in THz quantum cascade lasers by examination of temperature dependence of output power," Applied Physics Letters **106** (2015).
17. G. Liang, E. Dupont, S. Fathololoumi et al., "Planar integrated metasurfaces for highly-collimated terahertz quantum cascade lasers," Scientific reports **4**, 7083 (2014).
18. Y. Han, L. Li, J. Zhu et al., "Silver-based surface plasmon waveguide for terahertz quantum cascade lasers," Optics express **26**, 3814-3827 (2018).
19. T. Inoue, M. Yoshida, J. Gelleta et al., "General recipe to realize photonic-crystal surface-emitting lasers with 100-W-to-1-kW single-mode operation," Nature Communications **13**, 3262 (2022).
20. S. Noda, K. Kitamura, T. Okino et al., "Photonic-crystal surface-emitting lasers: Review and introduction of modulated-photonic crystals," IEEE Journal of Selected Topics in Quantum Electronics **23**, 1-7 (2017).
21. J. Ryu, C. Sigler, C. Boyle et al., "Surface-emitting quantum cascade lasers with 2nd-order metal/semiconductor gratings for high continuous-wave performance," in *Novel In-Plane Semiconductor Lasers XIX*(SPIE2020), pp. 189-199.
22. Y. Jin, L. Gao, J. Chen et al., "High power surface emitting terahertz laser with hybrid second-and fourth-order Bragg gratings," Nature communications **9**, 1407 (2018).
23. O. P. Marshall, V. Apostolopoulos, J. R. Freeman et al., "Surface-emitting photonic crystal terahertz quantum cascade lasers," Applied Physics Letters **93**, 171112 (2008).
24. M. Yokoyama, and S. Noda, "Finite-difference time-domain simulation of two-dimensional photonic crystal surface-emitting laser," Optics Express **13**, 2869-2880 (2005).
25. R. Colombelli, K. Srinivasan, M. Troccoli et al., "Quantum cascade surface-emitting photonic crystal laser," Science **302**, 1374-1377 (2003).

26. C. Peng, Y. Liang, K. Sakai et al., "Coupled-wave analysis for photonic-crystal surface-emitting lasers on air holes with arbitrary sidewalls," Optics Express **19**, 24672-24686 (2011).
27. Y. Liang, C. Peng, K. Sakai et al., "Three-dimensional coupled-wave model for square-lattice photonic crystal lasers with transverse electric polarization: A general approach," Physical Review B **84**, 195119 (2011).
28. Y. Liang, C. Peng, K. Ishizaki et al., "Three-dimensional coupled-wave analysis for triangular-lattice photonic-crystal surface-emitting lasers with transverse-electric polarization," Optics express **21**, 565-580 (2013).
29. Y. Yang, C. Peng, Y. Liang et al., "Three-dimensional coupled-wave theory for the guided mode resonance in photonic crystal slabs: TM-like polarization," Optics Letters **39**, 4498-4501 (2014).
30. Y.-C. Liu, and K. Chang, "Simple implementation of effective permittivity at dispersive metal-dielectric tilt interfaces for open-source FDTD package," in *2013 USNC-URSI Radio Science Meeting (Joint with AP-S Symposium)*(IEEE2013), pp. 75-75.
31. L. Sirigu, R. Terazzi, M. I. Amanti et al., "Terahertz quantum cascade lasers based on two-dimensional photonic crystal resonators," Optics Express **16**, 5206-5217 (2008).
32. Y. Chassagneux, R. Colombelli, W. Maineult et al., "Electrically pumped photonic-crystal terahertz lasers controlled by boundary conditions," Nature **457**, 174-178 (2009).
33. Y. Yang, C. Peng, and Z. Li, "Semi-analytical approach for guided mode resonance in high-index-contrast photonic crystal slab: TE polarization," Optics express **21**, 20588-20600 (2013).